# Generative AI in developing User Experience Research Point of View: A NotebookLM case study

**Mona Giff**

London, UK monaegiff@gmail.com

**Stephen Giff**

Google, USA, sgiff@google.com

**Huseyin Dogan**

School of Computing and Engineering, Bournemouth University Poole, UK, hdogan@bournemouth.ac.uk

## Abstract

User Experience Research (UXR) is currently undergoing a transition from traditional usability testing towards design-led and data-driven approaches, yet it faces an identity crisis due to a lack of methodological grounding in UXR and time-intensive methodologies which often lag behind product decision cycles. To address this, the UXR Point of View (PoV) framework formalises the UXR process by transitioning from raw data collection to forming an evidence-based PoV which drives strategic product impact. Furthermore, the use of GenAI in UXR has been investigated, but researchers often face increased work intensity when using GenAI, attributed to time spent on prompt engineering, data cleaning, and verification of AI outputs. This paper proposes and evaluates a formalised methodology for leveraging GenAI, specifically Google's NotebookLM, to augment the UXR PoV process. The methodology consists of five prompts across four stages: (1) leveraging the framework, (2) establishing roadmaps, (3) applying best-practices, and (4) crafting PoV narratives; and was tested on 11 UXR papers. Results showed that by using the proposed methodology, NotebookLM successfully leveraged the UXR PoV framework across all stages of PoV creation. These findings demonstrate that NotebookLM can serve as an effective collaborative partner in UXR, so long as it is provided with sufficient context and specific prompting.

## Background and Introduction

In the past decade, User experience research (UXR) has shifted from traditional usability testing toward design-led, quantitative, and data-driven approaches [1]. It has been argued that the discipline is currently facing an identity crisis due to several factors such as the democratisation of research, emerging disciplines, and time-intensive methodologies, which lead to product decisions being made before the research is complete [1]. In a systematic review of over 400 UXR studies from 2000-2016, it has been found that there is a perceived lack of methodological grounding in UX publications, including an absence of research questions or testable, articulated hypotheses, and a continued blurring of the lines usability and UX [2]. A recurring recommendation has been to adopt more rigorous standards for evaluating research, scrutinising the research methodologies, to turn insights from UXR into practices which can be implemented in an organisational context [1] [2].

Dogan, Giff, and Barsoum [3] have developed a structured framework designed to help UX researchers transition from raw data collection to forming an evidence-based Point of View (PoV) which drives strategic product impact. Their primary goal of this framework is to increase the validity of UXR by provided a flexible framework which can be established across various contexts. This framework is grounded in the theory of situational awareness (SA), as proposed by Endsley [4], operationalised as the perception of environmental elements, the comprehension of their meaning, and the projection of their future. The authors correlate a UXR PoV with the cognitive hierarchy of SA, a progression where data becomes information, information becomes knowledge (SA), and knowledge used to predict consequences leads to understanding, and ultimately wisdom. They view building a PoV as achieving this 'Wisdom'. Dogan et al. developed a four-level pyramid to represent the methodological progression required to build a PoV Furthermore, this framework is being used to develop a UXR 'Playbook'. The framework is leveraged to create a PoV, but also to create 'playbook cards', or individual cards which represent potential catalysts or obstacles. They include 'premeditated 'Best Practice' tactics, an 'be ready' tactics designed to thwart roadblocks. These ultimately can be used to create PoV narratives, which is a combination of cards aligned to a specific scenario, such as a 'Stakeholder Buy-in' narrative, which could include cards for cross-functional collaboration and demonstrating impact. This model therefore formalises UXR by providing a flexible framework, which can be applied to all UXR. Moreover, the founders of UXR PoV research. [5] argue that generative artificial intelligence (GenAI) can be leveraged for the PoV framework and suggest four steps for this. These steps are used in the current paper and details are provided in the methods section.

The use of AI in UXR has become increasingly popular. In investigating the qualitative experiences of UX Researchers at IBM Software in integrating GenAI into their professional workflows, GenAI was seen as an 'embryonic moment' where GenAI had not yet met researchers' expectations regarding efficiency and quality [5]. A productivity paradox emerged, in which IBM researchers found that GenAI use often increases work intensity, attributed to the trade-off of the time spent on prompt engineering, data cleaning, and mandatory manual verification of AI outputs. Therefore, the state of using GenAI in UXR is still budding and has many issues. This suggests the need for a formalised GenAI process for UXR, which can be achieved through the UXR PoV framework.

More specifically, the integration of GenAI technologies into the existing UXR PoV framework has been investigated [6]. The authors suggest how GenAI can be used to augment, rather than replace, human research within the UXR PoV framework. They suggest the following best practices:

- Data preparation: Data must be cleaned and formatted correctly for AI models
- Prompt Engineering: Must provide clear, concise instructions and sufficient context for the model to understand the research goals
- Critical evaluation: Must critically review and validate AI outputs using alternative methods or expert review to ensure accuracy and reliability.

However, despite their suggestions, a formalised methodology for leveraging GenAI for the UXR PoV framework has yet to be developed.

NotebookLM is Google's AI research tool which can analyse large collections of information and handle several uploaded formats [7]. NotebookLM is an LLM which answers queries based on sources which are uploaded. It analyses the given inputs to form an understanding of the topic and answers the questions using only the uploaded sources. For this reason, it can be seen as essentially creating an 'expert' AI agent in whatever the topic the research is grounded in and be used to streamline or augment the researcher's process. It can be especially useful in literature reviews, as this is normally a time-intensive process. The current paper aims to propose a methodology for the use of NotebookLM in developing UXR PoVs, in line with the framework's four stages [5]. This methodology will also be used to create PoV cards and associated narratives, according to the UXR PoV "Playbook".

## Method

This section provides an overview of the study design including samples, materials and prompts utilised.

(1) *Design* -the current methodology is comprised of five prompts across the four suggested stages, which can be inputted into NotebookLM, either sequentially or all at once, which together leverage GenAI and the UXR PoV framework for PoV creation.
(2) *Sample* – 11 papers published in ArXiv used in UXR PoV workshop for CHI2025.
(3) *Materials* -PDFs of all papers were uploaded to NotebookLM (Gemini-3). In addition to these papers, supporting materials were also uploaded to provide context. This included:
    - UXR PoV Playbook website: Definitions/FAQs page
    - UXR PoV Playbook website: Playbook Details
    - Giff, Dogan, and Barsoum (2024) Paper "*User Experience Research: Point of View Playbook"*
    - Example Play card taken from website
    - PoV Template

    This process was completed on a Macbook Air M2 (2022), with MacOS Sonoma 14.61, using a Safari browser.

(4) *Prompts-* were developed in line with suggestions from past research [5] [6] [8] and with the four suggested stages from Dogan et al [9].The prompts were tested iterated on until they produced the desired outputs in NotebookLM. This produced the final prompts:

    Phase 1: Leverage GenAI and the UXR POV Framework:

    *Prompt 1: Imagine you are a UX researcher interested in turning data into actionable insights for product strategy and design, providing a focused perspective on user needs, pain points, and motivations (see "FAQs 2 - UXR POV Playbook" and "POV Playbook Details - UXR POV Playbook" and "chiea24-89+(3).pdf" for more background). You want to apply the UXR Framework to cases (i.e. 1+GenAI_POV_CHI2025_asanandaji-2.pdf) and eventually turn the insights from this case into actionable POVs and plays. To begin, only focusing on the PDF "1+GenAI_POV_CHI2025_asanandaji-2.pdf" (but using the information from the other sources to guide you), generate a research question and hypothesis, extract themes (format this as a table), and give a two-sentence summary.*

*Make sure these answers are based only on the sources I gave you. They should be in a PhD level academic tone.*

Phase 2: Establish a Foundational Plan/Roadmap:

*Prompt 2: using the same logic as the previous prompt (background, academic tone, using only given pdfs/sources), please identify the following within the given paper ("1+GenAI_POV_CHI2025_asanandaji-2.pdf"):*

- *Project goals*
- *Data sources*
- *User needs*
- *Stakeholder information (in order to facilitate stakeholder-buy in and alignment at the foundational level of the framework)*

*Remember, this is using the information given in the other sources in order to guide your response.*

*Prompt 3: using the same logic as the previous prompt (background, academic tone, using only given pdfs/sources), identify the top 5 key insights for user needs and stakeholder information from only the PDF ("1+GenAI_POV_CHI2025_asanandaji-2.pdf"). Be specific, as these will be used to create POV play cards and be applied to the Playbook's pyramid structure.*

Phase 3: Apply GenAI-Enhanced Best Practices:

*Prompt 4: Create a series of UXR POV Playcards*) *that can be leveraged for UXR practioners to* implement various tactics and methods for progressing toward establishing a point of view.

*(Using the key insights from the previous prompt, and public UX Guidelines*), *and the logic from all previous prompts (background, academic tone, using only given pdfs/sources) create POV "play cards". Use the source "Share+Success.jpg.webp" as a template. Remember, you are an academic researcher interested in leveraging UXR into actionable insights for product strategy and design, providing a focused perspective on user needs, pain points, and motivations.*

*Provide two versions of each card:*

1. *With guidelines that are specific to the research, insights, and point of view*
2. *With generalized guidelines that can be leveraged in other research contexts. This card should still be grounded within the context of the current paper but have more generalised applications.*

*Furthermore, if there are no related cards in our history of creating cards, do not include related cards. Instead, keep the related cards section, but leave it blank.*

Phase 4: Craft compelling POV Narratives:

*Prompt 5: Using the playcards from the previous prompt, and the logic from all previous prompts (background, academic tone, using only given pdfs/sources) craft POV narratives, as well as any supporting material you feel is necessary, all tailored to different stakeholders (only relevant - based on PDF "1+GenAI_POV_CHI2025_asanandaji-2.pdf"). Remember, you are an academic researcher interested in leveraging UXR into actionable insights for product strategy and design, providing a focused perspective on user needs, pain points, and motivations.*

It is important to highlight that one of the most salient suggestions from Sanandaji and Stegbauer [6], was to provide the AI agent with enough context. As such, the first prompt establishes context, taking the PoV of a UX researcher interested in the UXR PoV framework additionally referring to sources which explain the UXR PoV Framework. Lastly, the prompts specify that all outputs should be in an PHD-level academic tone. This ensures that the AI agent has enough context to complete the task at hand, while also ensuring that all responses are at an appropriate level for research.

In addition, NotebookLM is known for its 'hallucinations', in which the model generates content which does not exist in the source text [9]. Therefore, when pointing to a source in the prompts, NotebookLM was directed to only focus on this information (i.e does not make anything up). Furthermore, checks were employed randomly to ensure that the AI agent was only using information from the targeted sources. These checks were in the form of questions asking where the model had sourced their information. This was used periodically to ensure accuracy.

The uploaded sources were selected at the beginning of the process for each paper and were referenced in each prompt. Once the prompting was completed for one paper, the process re-started, changing the prompts to include the PDF name of the target paper. All prompting across all papers was done in one NotebookLM notebook.

**Results and Discussion**

All generated outputs answered each prompt appropriately at every stage and across the designated literature. The outputs and associated checks showed that NotebookLM not only used the targeted paper to shape responses, but also the supporting materials. This suggests that sufficient context was provided both in the prompts and the uploaded supporting material, for NotebookLM to create PoVs. All answers were in a PhD-level academic tone, as instructed. Furthermore, the hallucination checks across all papers showed that NotebookLM was responding based on the uploaded sources, suggesting that the prompts were specific enough in their wording to prevent hallucinations.

However, to discuss the outputs from the proposed methodology in depth, the outputs from paper "*From 600 Tools to 1 Console: A UX-Driven Transformation"* written by Smith, Meijor-Irons, and Millar (2025) [11] will be examined. This paper examined the consolidation of a fragmented internal information system by Google's Internal Infrastructure UX team into a single, unified console, as guided by UXR PoV. They leveraged the UXR PoV framework at every layer to address this issue, eventually creating a PoV which addresses the burden of cross-system debugging by prioritising tool consolidation and a shared design language. They used the framework to develop and execute plays, which allowed the team to secure executive buy-in through a high-impact ROI metric: the amount of software engineer capacity saved annually. This paper and associated outputs will be discussed at every phase.

<u>Phase 1: Leverage GenAI and the UXR POV Framework</u>

The aim of this phase was to use GenAI (NotebookLM) to streamline various stages of PoV creation, including automated theme extraction, summarisation of targeted papers, and hypothesis generation. As such prompt 1 instructed NotebookLM to generate a research question and hypothesis, extract themes, and summarise the research question of this paper.

The outputs all appropriately addressed the prompt's instructions. The full output is available in the appendix. The research question and hypothesis were appropriate in scope and content for this paper. However, because this was done on a single, pre-existing paper, rather than a body of work, the hypotheses and research questions, representative of the foundational stage of the framework, this was less of a generated research question and hypothesis but rather extracting it from the paper.

| Theme | Description |
| --- | --- |
| **Information Fragmentation** | The complexity arising from over 1,000 distinct internal tools with inconsistent UIs and documentation gaps. |
| **Cognitive Friction** | Usability challenges in niche, unsupported solutions that impede efficient workflow navigation and slow down users. |
| **Debuggability Crisis** | A core pain point representing 67% of open-ended feedback, where intersecting factors across disparate tools make troubleshooting "impossible". |
| **Metric-Driven ROI** | The strategic shift from descriptive dashboards to a singular success metric—SWE capacity saved—to align with executive objectives. |
| **Strategic Consolidation** | Transitioning from high-level ambiguity to a unified console vision (Sigma) through service blueprinting and story mapping. |

*Table 1 Extracted Themes from Target Paper*

The extracted themes were appropriate and were later used to inform the 'Play Cards' (see table 1). They were specific to the paper, but general enough to be applicable across different contexts, and upon cross-checking, were deemed representative of the content of the paper. Furthermore, the summary provided a brief, but effective summary of the work (see Appendix A). Though this stage of the framework would usually be done on a corpus of data, the outputs demonstrate that NotebookLM is able to extract this information from an existing paper.

Phase 2: Establish a Foundational Plan/Roadmap

The goal of this phase was to use NotebookLM to define project goals, user needs, data sources, and stakeholder information. Prompt 2 asked NotebookLM to identify this information within the current paper. The outputs accurately responded to the instructions. The general project goals aligned with the research question, highlighting the consolidation of internal infrastructure tools, reducing cognitive load, streamline developer workflows, and establishing an ROI. These also align with the previously extracted themes. Furthermore, NotebookLM correctly identified the various data sources, such as semi-structured user interviews, stakeholder interviews, and a survey (see Appendix A for link to full output). Checks ensured that this information was correctly sourced from the paper. User needs were also identified and aligned well with those identified within the investigation. For example, NotebookLM identified rapid and efficient debuggability across systems and reduced context-switching between disparate dashboards. Lastly, Google's executive leadership, the technical infrastructure team, engineering partners, and product management teams were the stakeholders within the investigation, as determined by NotebookLM.

Prompt 3 instructed NotebookLM to identify the top five key insights for user needs and stakeholder information. These were based on the themes extracted in phase 1. Not only this, but they align well with the actual identified components which the authors suggested could constitute potential plays. For example, NotebookLM identified Metric Selection as Stakeholder Buy in as one of the insights. This is operationalised as SWE hours saved. Within the original paper, the authors highlight the metric which they developed, SWE, to demonstrate return on investment. Therefore, the methodology was successful in extracting insights from the current paper.

Phase 3: Apply Gen-AI Enhanced Best Practices

The aim of phase 3 was to integrate NotebookLM into the UXR PoV pyramid structure by creating play cards. Therefore prompt 4 asked NotebookLM to create play cards using the previously identified key insights. It was prompted to create two versions of each card: (1) with guidelines which are specific to the paper and (2) with generalised guidelines. The full set of generated cards can be seen in Appendix A. Four cards were generated, showing that the methodology was successful for the creation of these cards. It should be noted, it is unclear why only four cards were produced, when five key insights were identified. It may be that ‘debuggability as foundational friction’ is not so much of a play as much as an identified issue. The specific cards were appropriate for the context and scope of the target paper. However, the generalised cards were not as appropriate as the specific ones. Specifically, they often were not grounded in the context of the original paper. To highlight this, both versions of the card “Systematic Root Cause Mapping” are shown in Table 2.

| Attribute | Version 1: Specific to Internal Infrastructure Research | Version 2: Generalized Research Context |
|---|---|---|
| **Issue** | Troubleshooting "effectively impossible" intersecting factors across fragmented tools. | Interpreting complex data and avoiding stakeholder interpretation bias. |
| **Type** | Offensive Strategy (Best Practice) — Tactics to transform workflow navigation into actionable insights. | Offensive Strategy (Best Practice) — Tactics to uncover unbiased meanings that empower users. |
| **Quote** | "Insight Generation outputs should answer the 'why' of the insight and be actionable". | "Information converted into situational awareness becomes knowledge". |
| **Related Cards** | --------------------- | -------------------- |
| **Best Practice Guidance** | Use Service Blueprints to map user interactions and pain points across various service layers. By visually linking specific debugging obstacles to distinct stages of the developer journey, the researcher identifies the root cause of systemic "toil". This transitions data into an actionable POV that recommends tool consolidation rather than superficial UI fixes. | Move up the pyramid by converting correlated data into knowledge of the situational environment. Utilize methodologies like story mapping and service models to visualize relationships between disparate touchpoints. Providing an impartial meaning of the data prevents cross-functional (XFN) stakeholders from applying their own biases to the research conclusions. |

*Table 2 Card 2: Systematic Root Cause Mapping*

While the version one of this card is very specific and appropriate to the paper, the generalised research context is somewhat vague. The specific card highlights the use of service blueprints, to visually represent obstacles within the developer journey, facilitating the identification of systematic issues. The general card highlights the use of visualisation, preventing cross-functional stakeholders from applying their own biases to the research conclusions. Though they both identify the use of visual methods, the applications and rationale behind them are different. One would expect that the rationale would be similar between the cards, as they should both be grounded within the research, but the applications would be slightly different. Therefore, though the current methodology was successful in producing specific cards, more work is needed to create appropriate generalised cards.

Phase 4: Craft compelling PoV Narratives

The goal of this final phase was to develop clear, structured PoV narratives and supporting materials tailored for different stakeholders and various mediums. Prompt 5 thus asked NotebookLM to craft PoV narratives. NotebookLM generated narratives for three user types:

Executive Leadership, Engineering Leadership, and Product Management and UX Design. The full outputs can be seen in Appendix A. The Engineering Leadership POV is used to demonstrate the generated narratives:

> **Engineering Leadership POV: Debuggability and Toil Reduction**
>
> **Stakeholder Focus:** Software Engineering (SWE) leads concerned with technical debt, system reliability, and troubleshooting efficiency**.**
>
> - **The Narrative:** Analysis of the Data Collection phase revealed that debuggability is the most critical technical bottleneck, accounting for 67% of all developer dissatisfaction. The current "effectively impossible" state of troubleshooting arises from the intersection of disparate factors across unintegrated platforms (e.g., quota, logs, and storage issues). Our POV dictates that we must move beyond providing "more information" and instead prioritize centralized observability. By implementing a "Needs Attention" listing of error messages and warnings within a unified interface, we reduce the time-to-resolution for complex pipeline failures. This engineering-led consolidation, validated through R.I.T.E. studies, replaces numerous clicks and extensive scrolling with a shared design language that standardizes mental models across Compute, Networking, and Storage.
> - **Supporting Material:** The Service Blueprint and Story Map**.** Cross-functional visualizations that map user interactions, pain points, and tool intersections across various service layers to identify the root causes of troubleshooting friction.

The narrative encapsulates the data collection phase, highlighting a 'Needs Attention' listing of error messages and warnings within an interface, reducing the time-to-resolution for pipeline failures. This is in line with the previously identified card. In addition, it suggests supporting material in the form of the Service Blueprint and Story Map for visualisations of user interactions, to troubleshoot friction. This is once again in line with the PoV play cards. Taken together, this suggests that the methodology was successful in creating these PoV narratives. It should be noted that these were not in the typical 'PoV' template [5], but when this template was provided it made the narratives lose their specificity. In addition, the generated narratives did not consider the business context enough, which is usually in the template. This may have not been clear in the paper. These issues are likely due to only one paper being used, and the corpus of data not being available for NotebookLM. Further work is needed to workshop this.

## Conclusions and Future Work

This paper showed that NotebookLM can be leveraged to create PoVs, according to the UXR PoV framework, using the proposed methodology. This is extremely important in an era in which trust in AI is low, especially in organisational contexts [6] [14].This paper supports the notion that AI should be used as a collaborator, rather than a replacement for the human researchers, ensuring human-centred outcomes. As such, further AI research which adopts this view is necessary to educate the public on how to use AI in an appropriate fashion, especially in a corporate environment.

This investigation also confirmed that the previously suggested best practices for AI were successfully employed within the proposed methodology. Sanandaji and Stegbauer suggested that all instructions should be clear and concise and provide sufficient context for the model to understand the research goals [6]. The current investigation demonstrated that providing enough context ensures that NotebookLM adopted a UXR perspective. This method consisted of a continued conversation with NotebookLM, rather than just singular prompting, which provided further context. Furthermore, despite NotebookLM's tendency to hallucinate information [9], hallucination 'checks' showed that the methodology was specific enough. This is important when considering industry mistrust in the outputs of AI, or the view that GenAI increases work intensity due to mandatory manual verification and time spent on prompt engineering to avoid hallucinations [5].

One major limitation identified in this investigation is the non-deterministic nature of AI. This means that GenAI does not simply produce a single, fixed answer for an input, but a broad spectrum of responses [10]. Therefore, a direct replication of the outputs of the current paper is not possible because the same prompts on the same paper may produce similar insights, but they will be worded differently, or importance may be placed on a slightly different insight. However, this is not a limitation of the current paper, but rather the state of AI.

The current paper only applied the methodology to literature from the CHI2025 UXR PoV workshop, meaning that they are all centred around the UXR PoV framework. This may have made it easier for NotebookLM to apply the framework. As such, future research must investigate non-UXR PoV papers. Furthermore, future research will also apply this framework to non-UX, as a good research framework should be able to be applied in any research context. In addition, this methodology was only applied to a single paper at a time. Because of this, NotebookLM had trouble with populating the UXR PoV narrative templates, and the first phase, regarding the current paper, was rendered less important. In related research, GenAI has been successful in leveraging a corpus of data for PoV creation [11]). The models used in this investigation (ChatGPT, Gemini, NotebookLM, Microsoft Copilot) were able to establish hypothesis and extract themes and leverage insights from the data for PoV creation, and provide associated narratives, in line with templates. Therefore, this methodology should be used on a corpus of data to ensure that the associated prompts still produce similar outputs. These investigations would use the same methodology with slightly different wording within the prompts to ensure that NotebookLM understands that it is working with a corpus of data rather than pre-existing literature. Furthermore, the outputs from the current investigation showed that NotebookLM can leverage insights from a single paper, but this can most likely be applied for systematic/literature review. AI has already been shown to be used in Thematic Analysis [12]. The strength of NotebookLM lies in its ability to make sense of a corpus of literature and synthesise the results. Therefore, future research will investigate whether GenAI can leverage insights from a corpus of literature and apply it to the UXR framework.

**Appendix A**

“From 600 Tools to 1 Console: A UX-Driven Transformation” Outputs

Cards and Summaries for all papers